\documentclass[letterpaper,10t]{scrartcl}

\usepackage[utf8]{inputenc}
\usepackage[T1]{fontenc}
\usepackage[english]{babel}
\usepackage[bottom=3.9cm,top=3.9cm,left=2.6cm,right=2.6cm]{geometry}

\usepackage{amsmath}
\usepackage{graphicx}
\usepackage[justification=RaggedRight]{subfig}
\usepackage{url}
\usepackage[mode=text]{siunitx}

\usepackage{pdflscape}

\usepackage{algorithm}
\usepackage{algorithmic}

\usepackage{pxfonts}

\usepackage{hyperref}
\hypersetup{
colorlinks=true,
pdftitle={SingTel},
pdfauthor={Thomas Holleczek},
pdfsubject={},
pdfkeywords={}
}

\newcommand{\sectionname}[1]{{Section#1}}

\title{Digital breadcrumbs: Detecting urban mobility patterns and transport mode choices from cellphone networks}
\author{Thomas Holleczek, Liang Yu, Joseph K.~Lee, Oliver Senn, \\Kristian Kloeckl, Carlo Ratti, Patrick Jaillet}

\begin{document}
\maketitle

\begin{abstract}

Many modern and growing cities are facing declines in public transport usage, with few efficient methods to explain why. In this article, we show that urban mobility patterns and transport mode choices can be derived from cellphone call detail records coupled with public transport data recorded from smart cards. Specifically, we present new data mining approaches to determine the spatial and temporal variability of public and private transportation usage and transport mode preferences across Singapore. Our results, which were validated by Singapore’s quadriennial Household Interview Travel Survey (HITS), revealed that there are 3.5 (HITS: 3.5 million) million and 4.3 (HITS: 4.4 million) million inter-district passengers by public and private transport, respectively. Along with classifying which transportation connections are weak or underserved, the analysis shows that the mode share of public transport use increases from \SI{38}{\percent} in the morning to \SI{44}{\percent} around mid-day and \SI{52}{\percent} in the evening.

\end{abstract}

\section{Introduction} \label{sec:introduction}
Securing public transportation ridership is critical for developing a sustainable urban future. However, existing systems for analyzing and identifying weaknesses in public transport connections face major limitations. In cities, origin-destination (OD) matrices---which measure the flow of people between different geographical regions---are often generated using household surveys and roadside monitoring, approaches which are time consuming, expensive, and lack spatial and temporal accuracy~\cite{Caceres2007}. Focus in more recent research has been drawn towards using cellphones to overcome the limitations mentioned above---using the cellphone fluctuations in GSM signal strength~\cite{Sohn2006,Anderson2006,Anderson2008} and location, speed, and acceleration estimates obtained through GPS~\cite{Liao2007,Zheng2010,Stenneth2011} to detect individual trips and categorize them into classes such as walking, cycling, public transport and driving a car. These approaches are indeed innovative and capture in detail individual travel behavior, but are limited by their sample sizes (e.g.~number of volunteers) and currently face difficulties scaling up. While these methods will continue to be valid sources of data and insight, there are emerging opportunities to analyze urban transportation networks using data mining approaches, specifically in using cellphone call detail records to detect spatio-temporal patterns in urban mobility and transport mode preferences.

As cities not only become denser but also more extensive, there is an increasing pressure to maintain the connectivity and accessibility of goods and services through public transportation networks and integrated urban design. However, despite operating highly effective public transportation systems, modern and growing cities such as Singapore are facing declines in public transport usage; local governments and city planners are often equipped with few, efficient and innovative tools to explain where and why such declines are occurring~\cite{Choi2008}. Given that cities are becoming increasingly digitized and with the pervasiveness of cellphones, new data-driven methods are surfacing to help understand and explain characteristics of urban mobility~\cite{White2002,Pan2006,Caceres2007,Calabrese2011,Frias-Martinez2012}. Unlike previous work, we derive the mode share of public and private transportation (including taxis) by examining multiple datasets, namely cellphone call detail records and a public transportation smart card system. In this article, we use 3.4 million cellphone users’ call detail records to derive the inter-district (55 districts) mobility of people in the dense, metropolitan city-state of Singapore. We then subtract the number of Singapore’s public transportation smart card system users from our calculated value of urban mobility to estimate the share of private transportation users between districts. By examining the spatial and temporal patterns of public and private transportation, we can determine to and from where people prefer to take public or private transit and furthermore classify the time periods and the districts that may be considered weakly connected, or underserved, by public transportation. Ultimately, we present a novel method of using cellphone data to map the mobility of people in cities and posit that its application can not only benefit urban transportation and planning efforts but also other domains focused on or affected by the movement of people across space. 


\section{Datasets} \label{sec:datasets}
We use cellphone call detail records and a public transport dataset to estimate the flows of people using public and private transport between the 55 administrative districts of Singapore.

\subsection{Cellphone dataset} \label{sec:datasets:cellphone}
The cellphone dataset consists of location data of \SI{3.4}{} million anonymized users of SingTel, Singapore's largest telecommunications company with a market share of \SI{45.3}{\percent}.
%
The data were recorded during a two-month period from mid-March to mid-May 2011. A call detail record (CDR) includes the location of the cell tower each cellphone connects to and is created by the billing system of SingTel in the case of following network events:
\begin{itemize}
\item a phone call was initiated or received (at the beginning and at the end of the call).
\item a short message was sent or received.
\item the cell phone user accessed the data network (for example, to open a website or retrieve emails).
\end{itemize}
There is no information about handovers between base stations during long phone calls. Logging in and out of the cellphone network is also not recorded.

We use the location of the base station a cellphone connects to as the location of a cellphone user and estimate the localization error as follows.
%

The average service radius of a base station is approximately $r=\SI{212}{\meter}$. Due to confidentiality of the data, the locations of the base stations were offset by a maximum of \SI{200}{\meter}. This leads to an average uncertainty radius of \SI{412}{\meter} for the location estimation of cellphones. However, the cellphone might not always connect to the closest base station (for example, if the closest base station is busy), which can further increase the localization error.
%
A study by Ferris et al.~\cite{Ferris2007} shows that the median localization error can be reduced to $\SI{128}{\meter}$ with the help of Gaussian processes when considering the received GSM signal strength as well. The received signal strength is, however, not included in our dataset.

The location of a cellphone is only recorded in the case of the network events listed above, which means that subscribers using their cellphone frequently can be tracked more precisely. To measure how frequently cellphone subscribers access the cellular network, we introduce the network inter-event time. Let $\mathbf{t} = [t_1, t_2, \ldots t_n]$ be the timestamps of the network events related to the cellphone subscriber $\alpha$. We then define the network inter-event time $t_\alpha$ of $\alpha$ as
\begin{equation}
t_\alpha = \frac{1}{n - 1} \sum_{i=2}^n (t_i - t_{i-1}).
\end{equation}
We determined the distribution of the inter-event time. The average inter-event time for the entire dataset is \SI{320}{\minute}. The first quartile of the inter-event time is $t_{25} = \SI{41}{\minute}$, the median is $t_{50} = \SI{114}{\minute}$ and third quartile is $t_{75} = \SI{406}{\minute}$.

\subsection{Public transport dataset} \label{sec:datasets:public}
The public transport dataset consists of trips made by \SI{4.4}{} million anonymized users of Singapore’s public transport system during a two-week period in April 2011. As mentioned earlier on, we consider taxi trips as private rather than public transport in this study. In Singapore, passengers use smart cards when getting on and off trains and buses \cite{ezlink}. The smart card system records the station and the time of departure and arrival for each trip. Single trips from the same passenger within a certain short period are merged as a combined trip which reflects the actual origin-destination pair. One minor limitation of the dataset is that only the trips between stations are visible, while most people have to walk a short distance to their final destinations. However, the density of bus stations in Singapore is very high especially when compared to the size of the 55 districts.

\section{Approach} \label{sec:approach}
The flow of people between different geographical regions $R=\{r_1, r_2, \ldots\}$ in a given time interval is commonly represented through a quadratic origin-destination (OD) matrix $A(t_s,t_e)$. $A_{i,k}(t_s,t_e)$ counts the number of trips from region $r_i$ to region $r_k$ that arrive in the time interval $[t_s,t_e]$. OD matrices can be aggregated and normalized over time, for example to get hold of the average number of trips on Mondays between 6 and 9 in the morning. We determine origin-destination matrices of Singapore on an hourly basis.

The flows of people between different geographical regions of Singapore can be partitioned into flows by public and by private transport, as people travel either by a private car (including taxis) or by the public transport system. That means, the overall mobility $A_\textnormal{singapore}$ in Singapore can be decomposed into the OD matrix $A_\textnormal{public}$ containing trips by public transport and the OD matrix $A_\textnormal{private}$ containing trips by private transport:
\begin{equation}
A_\textnormal{singapore} = A_\textnormal{public} + A_\textnormal{private}.
\end{equation}
We estimate the overall mobility $A^*_\textnormal{singapore}$ in Singapore by upscaling the OD matrix $A_\textnormal{singtel}$ determined from the cellphone dataset (see \sectionname{~\ref{sec:approach:singtel}}) and determine the accurate public transport OD matrix $A_\textnormal{public}$ from the public transport dataset (see \sectionname{~\ref{sec:approach:public}}). To do this, we extract individual trips from the cellphone and the public transport datasets. We then map the start and end points of all determined trips to the 55 administrative districts of Singapore \cite{singaporeDistricts} and group them into hourly OD matrices. Finally, subtracting $A_\textnormal{public}$ from the estimated overall mobility $A^*_\textnormal{singapore}$ yields an estimate of the private transport OD matrix $A^*_\textnormal{private}$ (see \sectionname{~\ref{sec:approach:private}}).

As every public trip is captured by the public transport smart card system, the public transport OD matrix represents actual numbers. This means that the accuracy of $A^*_\textnormal{private}$ only depends on the accuracy of the overall mobility $A^*_\textnormal{singapore}$ estimated from the cellphone dataset. More specifically, every trip that is not detected in the cellphone dataset directly contributes to the underestimation of the number of private trips.

At first sight, it might seem that private trips are more difficult to detect from the cellphone data as people driving a  car are much less likely to use their cellphones than while traveling on a train or a bus. However, our trip extraction algorithm is based on the recognition of origins and destinations of trips and only requires cellphones be used before and after, but not during trips.

\subsection{Overall OD matrices} \label{sec:approach:singtel}
We determine the mobility of SingTel customers $A_\textnormal{singtel}$ from the cellphone call detail records with a trip detection algorithm and estimate the overall mobility $A^*_\textnormal{singapore}$ by upscaling $A_\textnormal{singtel}$ to the entire population. The idea behind the trip detection algorithm is that call detail records of a particular subscriber accumulate at origins and destinations of trips in the form of clusters. The algorithm can be applied to the call detail records of all subscribers or a selected subset of them. However, both options come with a downside:
\begin{enumerate}
\item Short trips cannot be detected with a clustering algorithm when a cellphone is not used frequently and no call detail records occur at the destination of a trip. Therefore, considering all SingTel subscribers to determine $A_\textnormal{singtel}$ adds a bias towards long trips.
\item To increase the detection rate of (especially short) trips, we could simply consider only subscribers that use their cellphones very frequently. However, this creates a bias, too, as the travel behavior of very frequent cellphone users might not represent the one of average users.
\end{enumerate}
As there is no way to recover undetected trips, we go for the second option and only consider \SI{1.5}{} million very frequent subscribers, whose inter-event time is less than \SI{60}{\minute} (see \sectionname{~\ref{sec:approach:singtel:extraction}}). Next, the bias introduced by only considering very frequent cellphone users must be removed (see \sectionname{~\ref{sec:approach:singtel:correction}}). Finally, we estimate the overall mobility $A^*_\textnormal{singapore}$ in Singapore by upscaling $A_\textnormal{singtel}$ to the entire population of Singapore (see \sectionname{~\ref{sec:approach:singtel:upscaling}}).

\subsubsection{Trip extraction} \label{sec:approach:singtel:extraction}
As mentioned before, the idea behind our trip detection algorithm is that call detail records accumulate at origins and destinations of trips in the form of clusters. We use an approach inspired by Calabrese et al.~\cite{Calabrese2011} to detect these clusters: 
\begin{itemize}
\item Let $\mathbf{x}_\alpha = [\mathbf{x}^1_\alpha, \mathbf{x}^2_\alpha, \ldots, \mathbf{x}^n_\alpha]$ denote the sequence of recorded locations of the cellphone user $\alpha$.
\item Then, the consecutive locations from $\mathbf{x}^k_\alpha$ through to $\mathbf{x}^{k+s}_\alpha$ (with $1 < k < k+s < n$) can be combined if $\Vert \mathbf{x}^k_\alpha - \mathbf{x}^{i}_\alpha \Vert < \Delta d \, \forall i \in [k+1,k+s]$. We use the threshold value $\Delta d = \SI{2}{\kilo \meter}$, which is sufficiently high to filter out jumps among adjacent cell towers that do not happen due to trips of the cellphone user in Singapore. We define the virtual location $\mathbf{y}^p_\alpha$ as the centroid of the combined locations:
\begin{equation}
\mathbf{y}^p_\alpha = \frac{1}{s + 1} \sum_{i=k}^s \mathbf{x}^i_\alpha.
\end{equation}
Let $\mathbf{y}_\alpha = [\mathbf{y}^1_\alpha, \mathbf{y}^2_\alpha, \ldots, \mathbf{y}^{m}_\alpha]$ denote the sequence of virtual locations determined for $\alpha$.
\item These virtual location do not necessarily have to be origins or destinations of trips as call detail records may also occur cumulatively during a trip (for example, when writing texts on a train) or a short interruption of a trip (for example when refueling at a gas station or when waiting for a bus). We therefore define clusters as those virtual locations that can be considered as origins and destinations of trips. A virtual location $\mathbf{y}^k_\alpha$ is referred to as a cluster if and only if there are at least two call detail records associated with $\mathbf{y}^k_\alpha$ and the time $\alpha$ spends at $\mathbf{y}^k_\alpha$ exceeds the threshold value $\Delta t$, which we select as \SI{20}{\minute}. 
Let $\mathbf{z}_\alpha = [\mathbf{z}^1_\alpha, \mathbf{z}^2_\alpha, \ldots, \mathbf{z}^p_\alpha]$ denote the sequence of clusters of $\alpha$.
\item We then define a trip $\mathbf{s}^k_\alpha$ as the path between the two consecutive clusters $\mathbf{z}^k_\alpha$ and $\mathbf{z}^{k+1}_\alpha$. $\mathbf{z}^1_\alpha$ is the origin of the first trip of $\alpha$ in the recorded period, and $\mathbf{z}^p_\alpha$ is the destination of the last trip. All other clusters $\mathbf{z}^k_\alpha$ serve as both origins and destinations of trips.
\end{itemize}
To determine the origin-destination matrix $A_\textnormal{singtel}(t_s, t_e)$, we map the start and end points of all detected trips to the 55 administrative districts of Singapore and add only these trips to $A_\textnormal{singtel}(t_s,t_e)$ whose \emph{end time} is $\in [t_s, t_e]$.

\subsubsection{Correction} \label{sec:approach:singtel:correction}
Next, the bias introduced by only considering the travel behavior of very frequent cellphone users has to be removed from the origin-destination matrices. The Land Transport Authority (LTA) conducts a Household Interview Travel Survey (HITS) every four to five years to give transport planners and policy makers insights into the traveling behavior of residents. About one percent of all the households in Singapore are surveyed each time, with household members answering detailed questions about their trips \cite{Choi2008}. According to the HITS from 2008, the travel behavior in Singapore is mainly shaped by home or work related trips: the origin or the destination of \SI{78}{\percent} of all trips is either a home or a work location \cite{Choi2008}. We therefore determine the distribution of home and work locations for both very frequent cellphone users and all SingTel subscribers from their call detail records. A comparison between these distributions indicates the local preferences of very frequent cellphone users and helps us identify differences in the travel behavior.

\begin{figure}
\begin{center}
\subfloat[\label{fig:dist}Number of home and work locations of SingTel customers for all 55 administrative districts of Singapore.]{\includegraphics[width=0.85\textwidth]{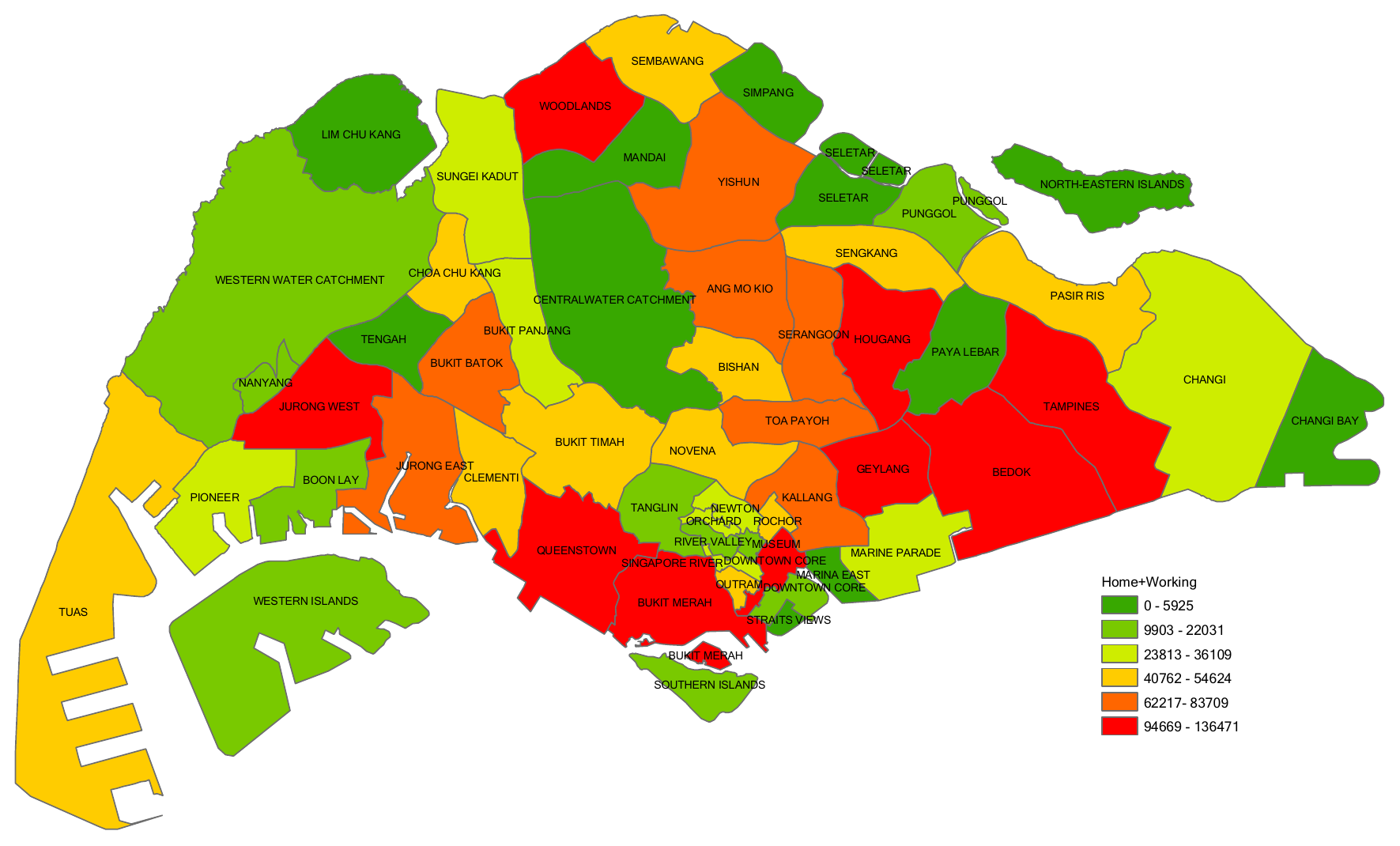}}\hfill
\subfloat[\label{fig:frequent}Share of very frequent cellphone users for residential districts of Singapore. The overall share of very frequent users is $\varphi = 0.34$. The highest shares are found in Nanyang (district with a lot of students) as well as the Downtown Core.]{\includegraphics[width=0.85\textwidth]{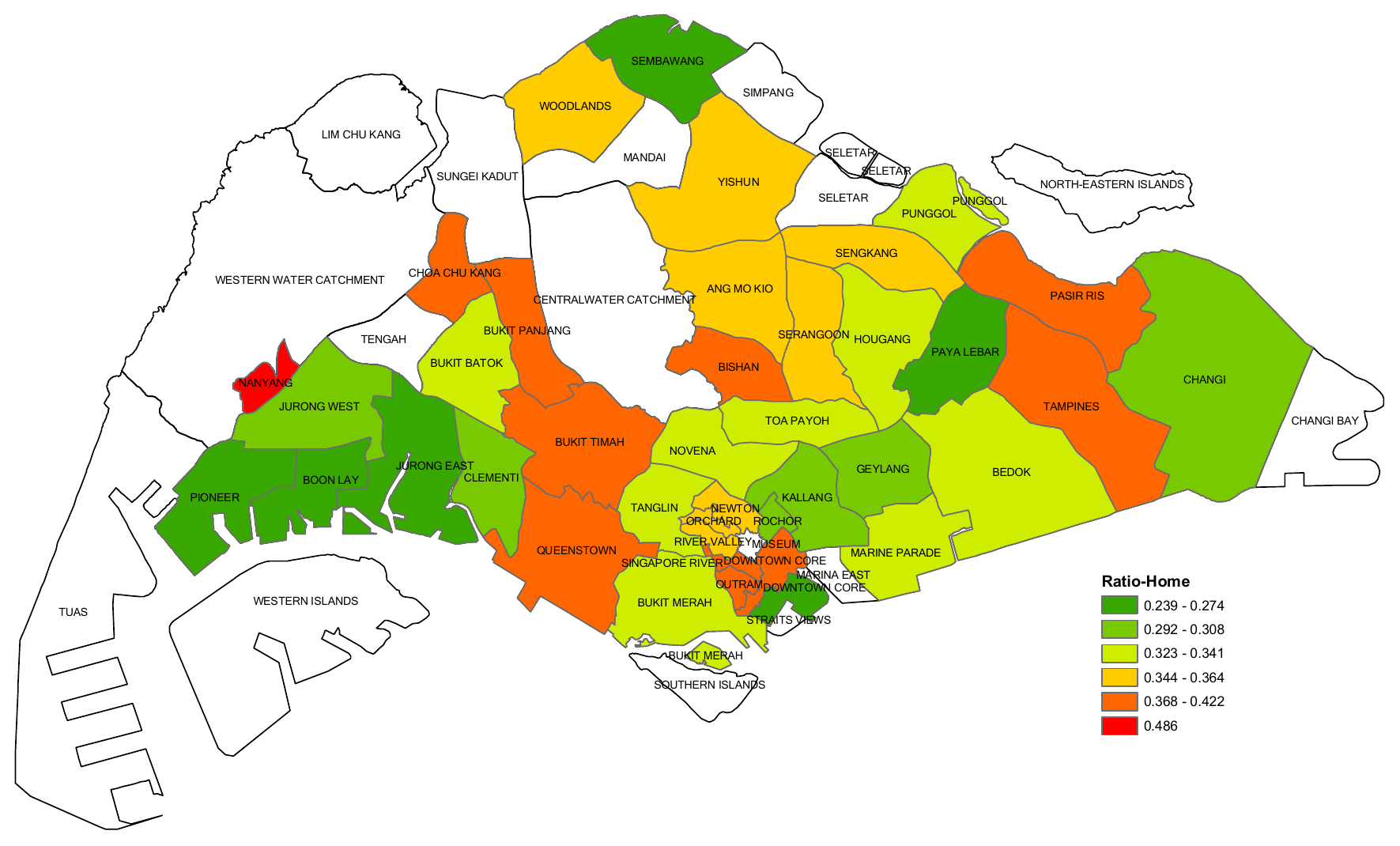}}
\end{center}
\caption{Number of home and work locations per district (a) and share of very frequent cellphone users (b).}
\label{fig:districts}
\end{figure}

Call detail records accumulate at home and work locations as these are the places where people spend most of their time. We therefore apply a slightly modified version of the K-means clustering algorithm to the recorded locations of the call detail records of each cellphone subscriber to determine their home and work locations:
\begin{enumerate}
\item Locations are only assigned to the closest centroid if the distance is less than the cluster radius $r = \SI{1.0}{\kilo \meter}$. Call detail records that are farther away than \SI{1.0}{\kilo \meter} from the closest centroid remain unassigned.
\item We are only interested in large clusters, which is why we ignore clusters with a share less than $x=0.15$. That is, at least \SI{15}{\percent} of all call detail records of a cellphone user have to be fused to form a cluster.
\end{enumerate}
We consider the identified clusters to be either a home or work location. The distribution of home and work locations of all SingTel subscribers is presented in \figurename{~\ref{fig:dist}}.


To remove the bias introduced by only considering very frequent cellphone users, we
correct the number of people traveling between two districts $i$ and $k$ as follows. We take into account the deviation of the share of the very frequent cellphone users living or working these two districts from the overall share. Let $\varphi_i$ denote the share of very frequent cellphone users in district $i$:
\begin{equation}
\varphi_i = \frac{m_i}{n_i},
\end{equation}
where $m_i$ is the number of home and work locations of very frequent users in district $i$, and $n_i$ the number of home and work locations of all subscribers in district $i$. \figurename{~\ref{fig:frequent}} shows $\varphi_i$ for residential districts of Singapore.
Moreover, let $\varphi$ denote the overall share of frequent cellphone users
\begin{equation}
\varphi = \frac{\sum_i m_i}{\sum_i n_i}.	
\end{equation}
We measured the overall share of very frequent cellphone users as $\varphi = 0.34$.
Then, $A_{ik}$, the number of people traveling from district $i$ to district $k$, can be corrected by taking into account the ratios $\varphi_i$ and $\varphi_k$ of very frequent cellphone users in these two districts as well as the overall share $\varphi$ as
\begin{equation}
A_{ik}' = A_{ik} \sqrt{\frac{\varphi^2}{\varphi_i \varphi_k}}.
\end{equation}
We use the square root to ensure the correlation between $A_{ik}'$ and $A_{ik}$ is linear regarding $\varphi$. The corrected OD matrix determined from the cellphone dataset is referred to as $A'_\textnormal{singtel}$.

\subsubsection{Upscaling} \label{sec:approach:singtel:upscaling}
Finally, the corrected OD matrix $A'_\textnormal{singtel}$ must be upscaled to represent the overall mobility of Singapore. Taking into account the market share of SingTel, we estimate there were about \SI{7.5}{} million cellphone subscribers in Singapore in 2011:
\begin{equation}
\textnormal{number of cellphone subscribers} = \frac{\textnormal{number of SingTel subscribers}}{\textnormal{SingTel market share}} = \SI{7.5e6}{}.
\end{equation}
A population of around \SI{5.2}{} million in 2011 \cite{StatisticsSingapore2012} yields a cellphone penetration of about \SI{144}{\percent} when not considering prepaid cellphones.

To estimate the overall mobility $A^*_\textnormal{singspore}$, we upscale $A'_\textnormal{singtel}$ by dividing it by the SingTel market share, the cellphone penetration in Singapore and the share of very frequent cellphone users:
\begin{equation}
A^*_\textnormal{singapore} = \frac{A'_\textnormal{singtel}}{\textnormal{market share} \times \textnormal{cellphone penetration} \times \textnormal{share of very frequent users}}.
\end{equation}
This scaling assumes that the travel behavior of SingTel customers represents the travel behavior of all people in Singapore.

\subsection{Public transport OD matrices} \label{sec:approach:public}
Trips can be extracted fairly easily from the public transport dataset as each entry represents a single trip. 
Should the time between two consecutive trips of a particular person be less than \SI{45}{\minute}---which is the time allowed by the smart card system to transfer \cite{SBS}---we consider this as transit time and combine these trips as we are only interested in the final destination of trips. Finally, the origins and the destinations of the trips are mapped to the 55 administrative districts of Singapore, and hourly public OD matrices can be determined. As each trip is captured by the smart card system, the resulting OD matrices are not an estimate but represent actual numbers of public transport trips.

\subsection{Private transport OD matrices} \label{sec:approach:private}
Subtracting the public transport OD matrix $A_\textnormal{public}$ from the estimated overall mobility $A^*_\textnormal{singapore}$ yields an estimate of the number of people that do not take public transport:
\begin{equation}
A^*_\textnormal{private} = A^*_\textnormal{singapore} - A_\textnormal{public}.
\end{equation}
As a trip is either public or private, the resulting OD matrix describes the flows of passengers using private transport.

\section{Evaluation} \label{sec:evaluation}
We use the results of the latest Household Interview Travel Survey (HITS) from 2008 to evaluate the performance of our trip extraction algorithm and the precision of the corresponding OD matrices. According to the HITS, \SI{20}{\percent} of all trips remain in the same district. On the contrary, only \SI{4}{\percent} of all trips remain in the same district according to our study. Short intra-district trips cannot be recognized from the cellphone dataset, which has the following two reasons:
\begin{enumerate}
\item In our study, we estimate the location of a cellphone as the location of the connecting base station. To filter out jumps between adjacent base stations, our trip recognition algorithm is based on a clustering algorithm. As a consequence, trips shorter than \SI{2.0}{\kilo \meter} cannot be detected with our dataset. This is slightly more than the expected distance of intra-district trips, which can be estimated as \SI{1.9}{\kilo \meter} as follows:

Singapore's 55 administrative districts cover an area of \SI{710}{\kilo \meter \squared}, which yields an average side length $s = \SI{3.6}{\kilo \meter}$ per district when assuming a squared shape for districts. According to \cite{Dunbar1997}, the average distance $d$ between two randomly selected points inside a square with the side length $s$ is
\begin{equation}
d = \frac{\sqrt{2} \bigl[2 + 5 \sqrt{5} + \ln{(\sqrt{2} + 1)} + 2\sqrt{2}\bigr	]}{30} s.
\end{equation}
Thus, the average distance of intra-district trips is \SI{1.9}{\kilo \meter}.
\item If two consecutive trips are too short, it is unlikely even for very frequent users with an inter-event time of \SI{60}{\minute} that a network event occurs, resulting in only one detected trip rather than two.
\end{enumerate}
%
We therefore do not consider intra-district trips in our analyses. The number of samples of the latest HITS is not sufficient to compare the number of people traveling on specific connections (only one percent of all households were interviewed), which is why we focus on the overall number of trips and the mode share of public transport. Our mobility study shows a good correspondence with the latest HITS and a more recent study by the Land Transport Authority:
\begin{itemize}
\item We estimate the number of inter-district trips (both public and private) as \SI{7.8}{} million per day. According to the HITS from 2008, there were \SI{9.9}{} million trips per day in Singapore, \SI{7.9}{} million of which were inter-district \cite{Choi2008}.
\item According to our study, \SI{45}{\percent} of the trips in Singapore are produced by public transport and \SI{55}{\percent} by private transport on average. These results are backed up by a travel survey conducted by the Land Transport Authority in 2011 (\SI{44}{\percent} by public transport and \SI{56}{\percent} by private transport)~\cite{LTA2011}.
\end{itemize}
These figures suggest that our trip detection algorithm for call detail records as well as the presented correction and upscaling methods produce good estimates of the overall mobility in Singapore as well as the flows of private transport.

\section{Results} \label{sec:results}
To identify underserved public transport connections depending on the time of the day, we determine aggregated OD matrices for the morning (6 am to 10 am), mid-day (10 am to 5 pm) and the evening (5 pm to 10 pm) based on the hourly OD matrices. In our analysis, we limit ourselves to workdays. We investigate the evolution of the public transport mode share over the day and identify underserved public transport connections.

\subsection{Mode share of public transport} \label{sec:results:share}
The mode share of public transport in Singapore increases with time of the day. It is \SI{38}{\percent} in the morning, reaches \SI{44}{\percent} around mid-day and peaks at \SI{52}{\percent} in the evening. One reason for this observation could be people feeling tempted to go to work by taxi when being in a hurry in the morning and using public transport back home in the evening, or parents dropping their kids at school on the way to work.

\subsection{Underserved public transport connections} \label{sec:results:weak}
\begin{figure}
\begin{center}
\subfloat[\label{fig:underserved:morning}Morning (6 am to 10 am).]{\includegraphics[width=0.62\textwidth]{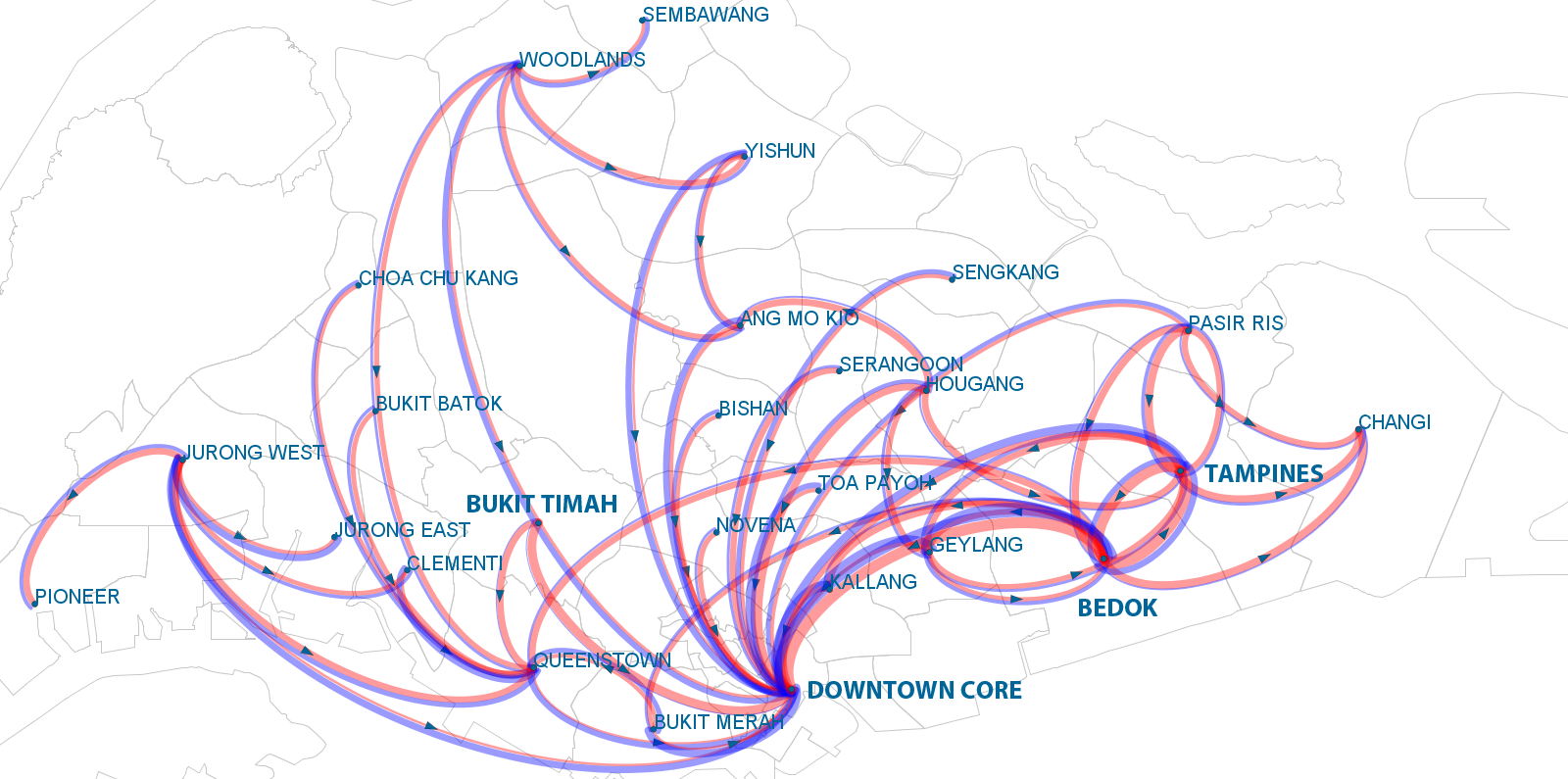}}\hfill
\subfloat[\label{fig:underserved:daytime}Mid-day (10 am to 5 pm).]{\includegraphics[width=0.62\textwidth]{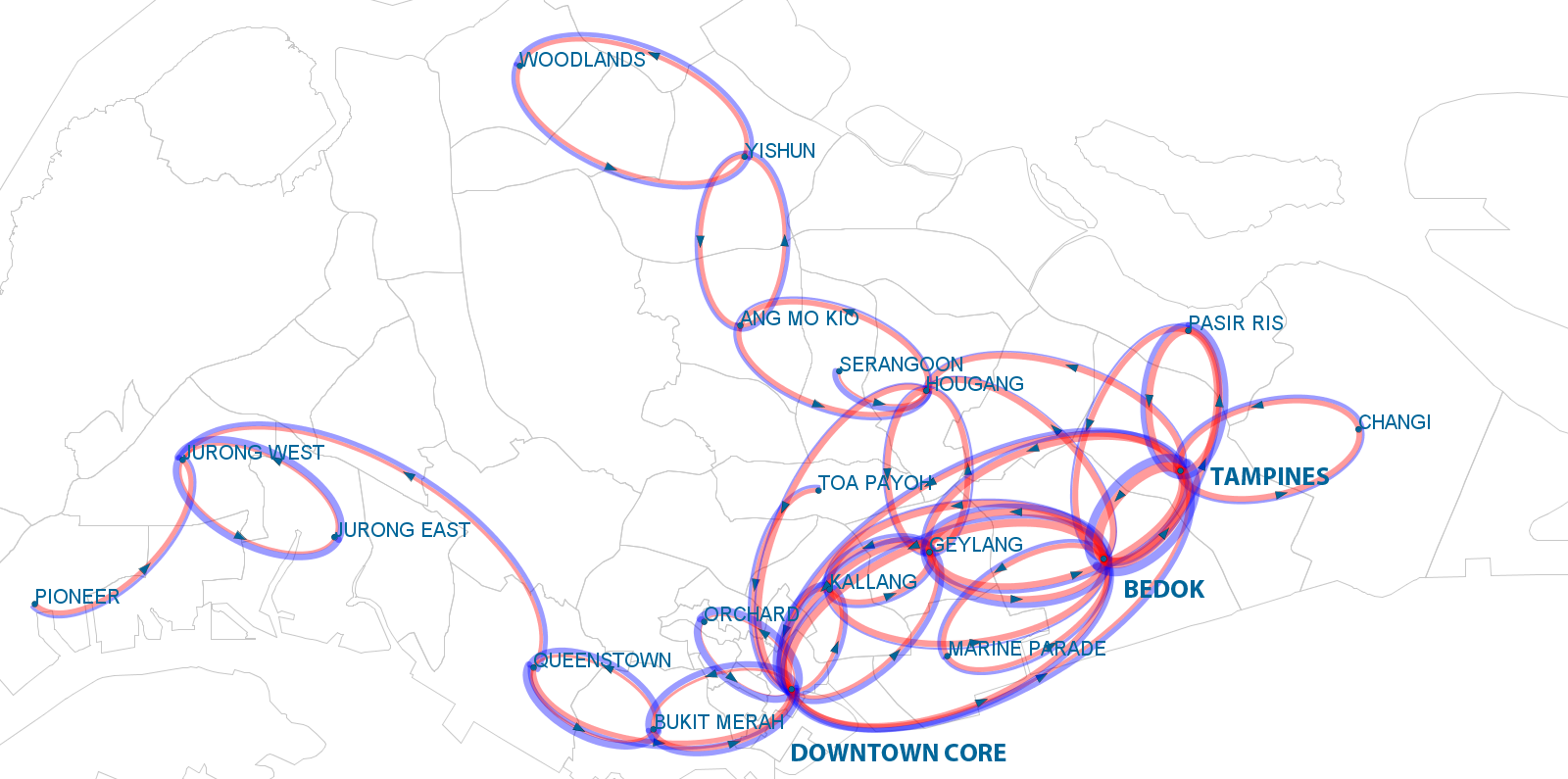}} \hfill
\subfloat[\label{fig:underserved:evening}Evening (5 pm to 10 pm).]{\includegraphics[width=0.62\textwidth]{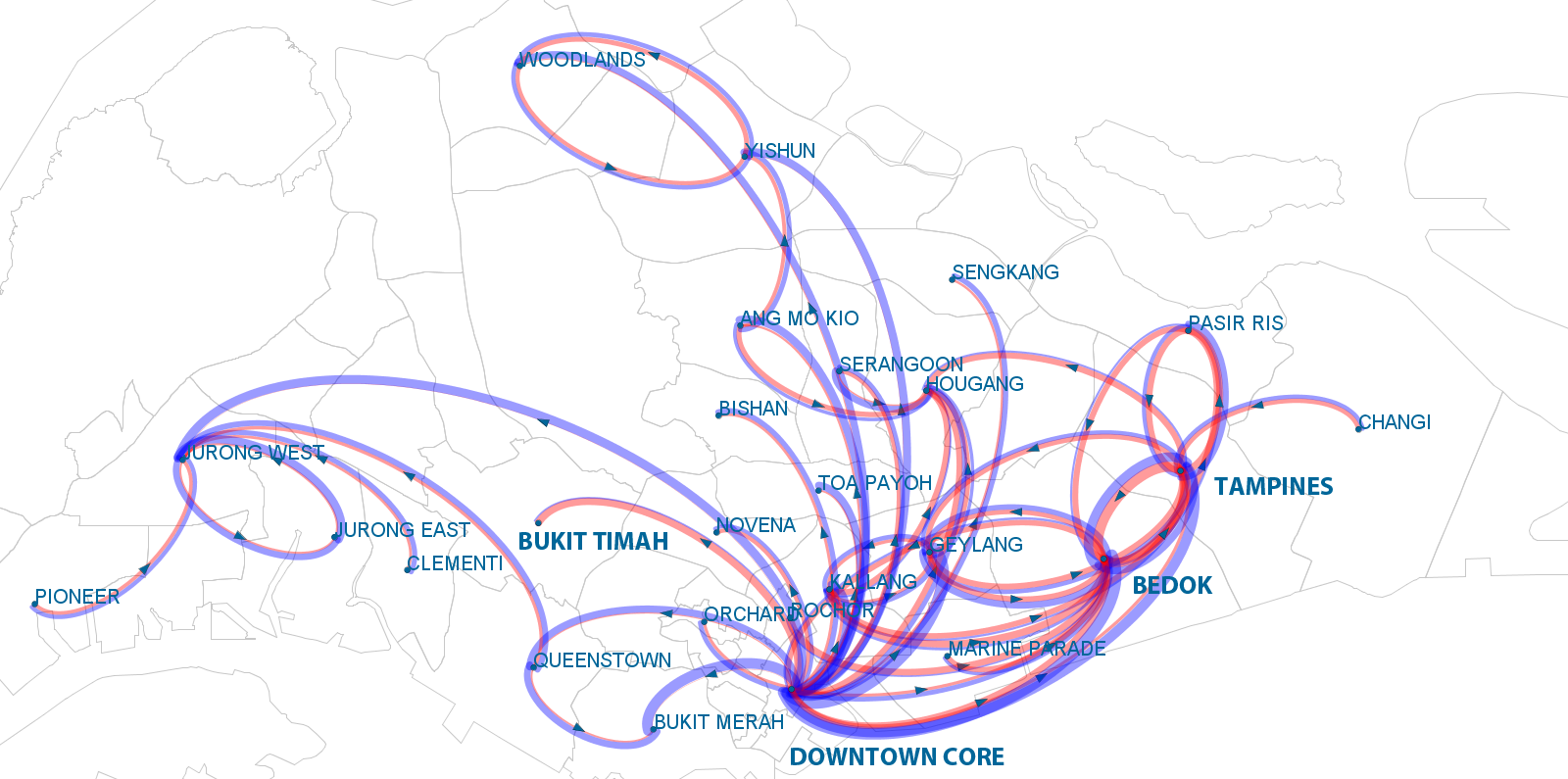}}
\end{center}
\caption{The 50 busiest inter-district connections, depending on the time of the day. The width of the lines represents the number of people traveling between the connected districts. Blue lines indicate the number of people traveling public transport, and red lines the number of people taking private transport. The highest mode share of private traffic can be observed between Bukit Timah and the city center in the morning and in the evening. Moreover, many underserved public transport connections can be found around Bedok and Tampines.}
\label{fig:underserved}
\end{figure}

We define a connection to be underserved by public transport when more people use private than public transport on this connection in a given period of time. Being particularly interested in the major connections, we present only the 50 busiest out of the 2970 inter-district connections in \figurename{~\ref{fig:underserved}}. The plots show the number of passengers taking public as well as private transport. Almost all trips are directed towards the city center in the morning. In the evening, the situation is reverse with most trips departing from the city center. The highest mode share of private traffic (around \SI{80}{\percent}) can be observed between Bukit Timah, a large residential area without access to a subway line, and the city center in the morning and in the evening. The Downtown Line, a new subway line in Singapore opening from 2013 to 2015, will cover this connection. Moreover, many underserved public transport connections can be found around Bedok and Tampines, the most and third-most populated districts, which are located in the east of Singapore.

\section{Outlook} \label{sec:outlook}
Traditional approaches to understanding public and private transportation flows through a city use travel surveys that are not only expensive, but also time consuming, inaccurate, and only sample a small percentage of a city’s total population. Using a data mining approach, we present methods and analyses that show that mobility and connectivity in cities can be accurately described by cellphone call detail records quickly, efficiently, in real-time, and with district-level spatial resolution or better. Coupled with public transport smart card records, we show that we can further identify the spatio-temporal variability of public and private transport use and begin to examine how and why such patterns exist. However, the value and reliability of such an approach is not limited to only this use-case. In fact, the implications of harnessing these types of datasets, specifically those being created from pervasive urban sensor networks and smartphones, are immense for planning and designing more livable and sustainable cities.

By analyzing these highly granular (individual-scale) datasets, produced in real-time, there are emerging opportunities to address a wide range of environmental, epidemiological, and socio-cultural questions that result from urban living and vis-à-vis human mobility in cities. With respect to transportation, using cellphone based mobility patterns opens up new ways to plan for the ever growing population of aging citizens---a particular issue for countries in Europe and in Japan, the United States, Singapore, among others. In order to guarantee accessibility to goods and services for populations with limited mobility options, big, urban datasets such as cellphone call detail records can help inform short-, medium-, and long-term decisions to plan, locate, and design, for example, mixed-use settlements, walkable connections and corridors, and mobility on-demand services. In a similar vein, examining mobility patterns and mode preferences and transfers as detected by cellphone call detail records and other transportation datasets can help to address common “last mile” problems that are typically difficult to detect and also discouraging for public transport use.

Lastly, a role for cellphone based urban mobility detection has significant implications for disease control in cities. As humans serve as the primary and secondary vectors of many infectious diseases, understanding from where people arrive and depart and by which transportation modes people are traveling, we have the potential to model how and where diseases might be spreading and from where they might originate. Big data analytics such as the analysis presented here, may introduce a distinctly new and interdisciplinary approach to modern epidemiological studies in both the developed and developing world.

Examining the “digital breadcrumbs” left behind by people in cities introduces new methods for examining urban mobility. Here, we emphasize the promising role for cellphone network analysis to generate meaningful descriptions of city scale transportation use and to comment on future avenues of urban research and planning.

\section*{Acknowledgements}
The authors would like to thank Clio Andris for processing the results of the Household Interview Travel Survey (HITS). Thanks to the National Science Foundation, the AT\&T Foundation, the MIT SMART program, the MIT CCES program, Audi Volkswagen, BBVA, The Coca Cola Company, Ericsson, Expo 2015, Ferrovial, GE and all the members of the MIT Senseable City Lab Consortium for supporting the research.

\phantomsection    
\addcontentsline{toc}{section}{References}
\bibliographystyle{unsrt}
\bibliography{/Users/tommy/Documents/MIT/myPublications/bibtex/senseablecitylab.bib}

\end{document}